\documentclass[conference]{IEEEtran}
\IEEEoverridecommandlockouts
\pdfoutput=1
\usepackage{cite}
\usepackage{amsmath,amssymb,amsfonts}
\usepackage{algorithmic}
\usepackage{graphicx}
\usepackage{textcomp}
\usepackage{xcolor}
\usepackage{multirow}
\usepackage{threeparttable}
\usepackage{booktabs}
\usepackage{epstopdf}
\usepackage[T1]{fontenc}
\usepackage{aecompl}
\usepackage{hyperref}
\usepackage{cleveref}

\def\BibTeX{{\rm B\kern-.05em{\sc i\kern-.025em b}\kern-.08em
    T\kern-.1667em\lower.7ex\hbox{E}\kern-.125emX}}
\begin{document}

\title{Invariant learning based multi-stage identification for Lithium-ion battery performance degradation\\
}

\author{\IEEEauthorblockN{Yan Qin}
\IEEEauthorblockA{\textit{Engineering Product Development} \\
\textit{Singapore University of} \\
\textit{Technology and Design} \\
Singapore, 487372 \\
yan$\_$qin@sutd.edu.sg}
\and
\IEEEauthorblockN{Chau Yuen}
\IEEEauthorblockA{\textit{Engineering Product Development} \\
\textit{Singapore University of} \\
\textit{Technology and Design} \\
Singapore, 487372 \\
yuenchau@sutd.edu.sg}
\and
\IEEEauthorblockN{Stefan Adams}
\IEEEauthorblockA{\textit{Materials Science and Engineering} \\
\textit{National University of Singapore} \\
Singapore, 117575 \\
mseasn@nus.edu.sg}}

\maketitle
\begin{abstract}
By informing accurate performance (e.g., capacity), health state management plays a significant role in safeguarding battery and its powered system. While most current approaches are primary based on data-driven methods, lacking in-depth analysis of battery performance degradation mechanism may discount their performances. To fill in the research gap about data-driven battery performance degradation analysis, an invariant-learning based method is proposed to investigate whether the battery performance degradation follows a fixed behavior. First, to unfold the hidden dynamics of cycling battery data, measurements are reconstructed in phase subspace. Next, a novel multi-stage division strategy is put forward to judge the existent of multiple degradation behaviors. Then the whole aging procedure is sequentially divided into several segments, among which cycling data with consistent degradation speed are assigned in the same stage. Simulations on a well-know benchmark verify the efficacy of the proposed multi-stages identification strategy. The proposed method not only enables insights into degradation mechanism from data perspective, but also will be helpful to related topics, such as stage of health.
\end{abstract}


\section{Introduction}
Ever-increasing stresses on environment protection and energy supply raise growing spotlights on Lithium-ion battery (LiB), which is a kind of renewable energy storage and power source in comparison with traditional fuel. Characterized by high-energy density and falling cost [1], wide applications of LiB have been observed in various electronics and instruments. Due to close connections between the health status of LiB and its powered systems, accurate evaluation and in-depth understanding of performance degradation become indispensable to ensure and improve the performance of the whole system.

Resulting from performance degradation, battery capacity is adopted as an indicator to indicate battery health state via available amount, which will fade due to battery aging. To distinguish from the capacity loss in calendar storage, in this article, capacity fade only refers to the ``cycle life'' loss in battery discharge ability over time. Related with complex physical and chemical mechanisms, performance degradation is a rather complex procedure. Besides, frequent usage and exposure to harsh environmental conditions will disturb the degradation speed. However, most literature on health state simply assume a single degradation behavior [2]-[4], which may not meet well with the real case.

Some literature pointed out the erratic behavior during capacity fading [5]-[8]. Spotnitz [5] found that the rate of capacity loss experiences initially high, then slows quickly, slows again, and finally rapidly decreases. Severson et al. [6] proposed a battery life prediction method before capacity degradation, indicating the recognition of different stages within a whole degradation procedure. To be more quantitative, Dubarry et al. [7] attempted to investigate fading stages through Peukert’s coefficient. Furthermore, Yu [8] put forward a degradation stage identification method using a distribution-based correlation index, which measured the similarity between the historic Gaussian components and the new Gaussian components. In this way, a whole degradation stage was classified into a health state, slight degradation state, and severe degradation state by checking the value of the defined index. However, selecting thresholds for these stages is manual and subjective, indicating the uncertainty of the results. Although these methods fail to give accurate identification results, they succeed in confirming the phenomenon of different degradation behavior. This phenomenon is named as multi-stage characteristic here. That is, a complete capacity fading may consist of several degradation behavior patterns, in which dynamics keeps consistent in the same pattern, while is dissimilar between different patterns. The consecutive discharging cycles in the same pattern form a degradation stage. The multi-stage identification will not only improve in-depth analysis of degradation mechanisms but also may contribute to other related topics, such as state of health estimation.

In this article, an invariant-learning based sequential stage division method is proposed, which is capable of providing reliable and unique results. The invariant is designed with the characteristics that it is statistically consistent in the same stage while it is different across stages. That is, they are sensitive to the changes in degradation speed. As a result, the identification of different stages can be iteratively achieved by monitoring the changes of the invariants. The stationary subspace analysis (SSA) method, which is good at learning stationary components from non-stationary systems, is improved to extract this kind of invariants with equally arranged cycling data.

The remaining part of this article is organized as follows. Section II gives a brief description of the discharging process and its associated dataset. The specifics of the proposed method are given in Section III. Then, the multi-stage division results are illustrated and discussed in Section IV. The conclusions are drawn in the last section.

\section{Process Description and Data Preparation}
Through repetitively charging and discharging procedure, energy is continuously transferred from LiB to its powered system. The repetitive usage over cycles causes a natural drop of capacity as shown in Fig. 1, in which data are provided by Ames Prognostics Center of Excellence in National Aeronautics and Space Administration (NASA) [9]. Capacity is the ability to evaluate available current that a battery can supply during a discharging cycle, which is usually recognized as the external performance indicator for LiB. In practice, for the $c$th discharging cycle, a two-dimensional data array $\mathbf X_c(J \times K_c)$ could be constructed with $J$ variables sampled over $K_c$ sample points. $K_c$ usually decreases with the increase of cycles in response to the capacity loss. As a result, a three-dimensional data matrix $\mathbf X(N \times J \times K_c)$ with uneven length is formed when the number of $N$ discharging cycles are available as shown in Fig. 2.
\begin{figure}[htb]
\centering
\includegraphics[scale=0.5]{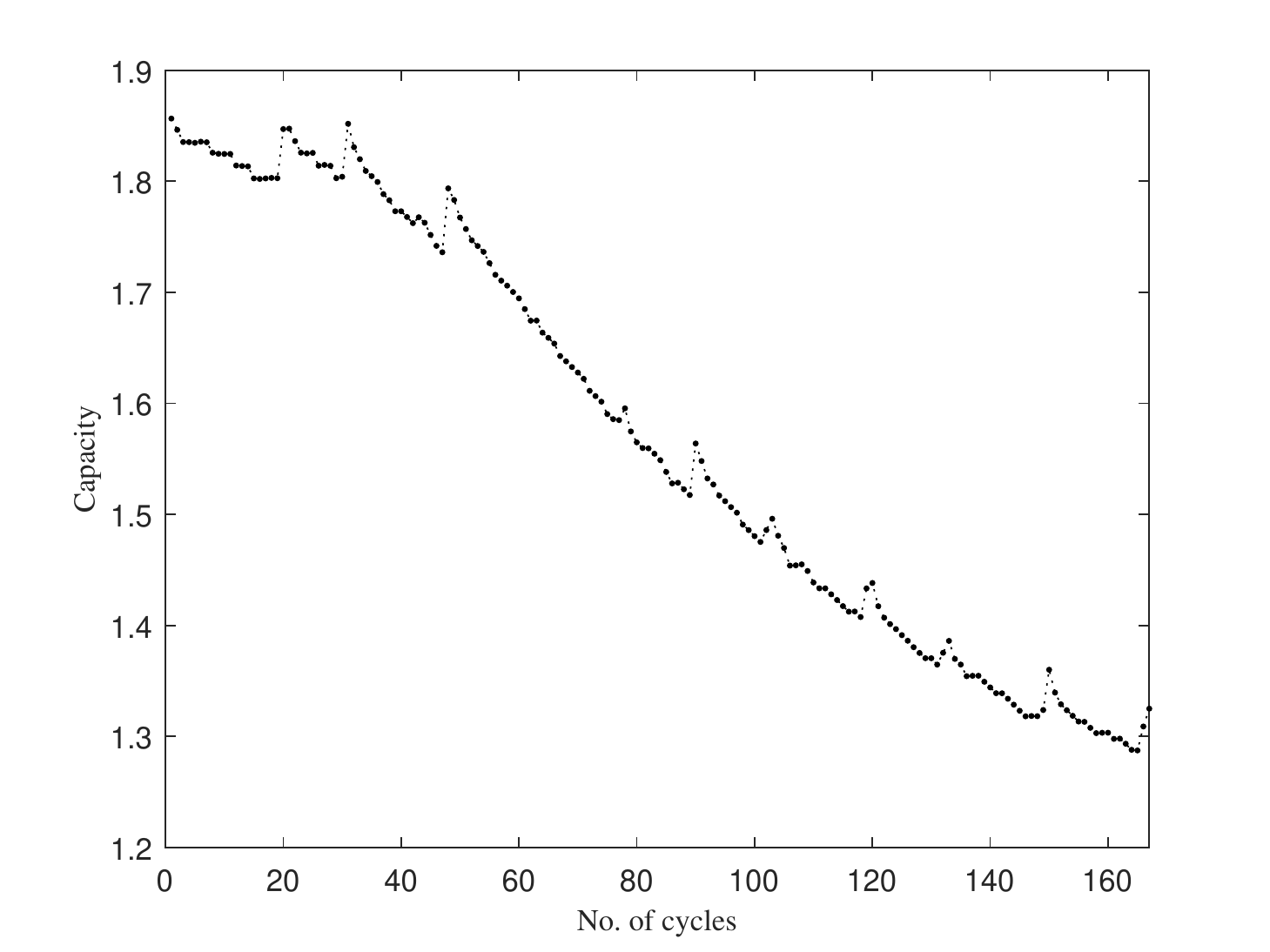}
\caption{The capacity loss over cycles of Battery B0005 provided by NASA.}
\label{MyFig1}
\end{figure}
\begin{figure}[htb]
\centering
\includegraphics[scale=0.4]{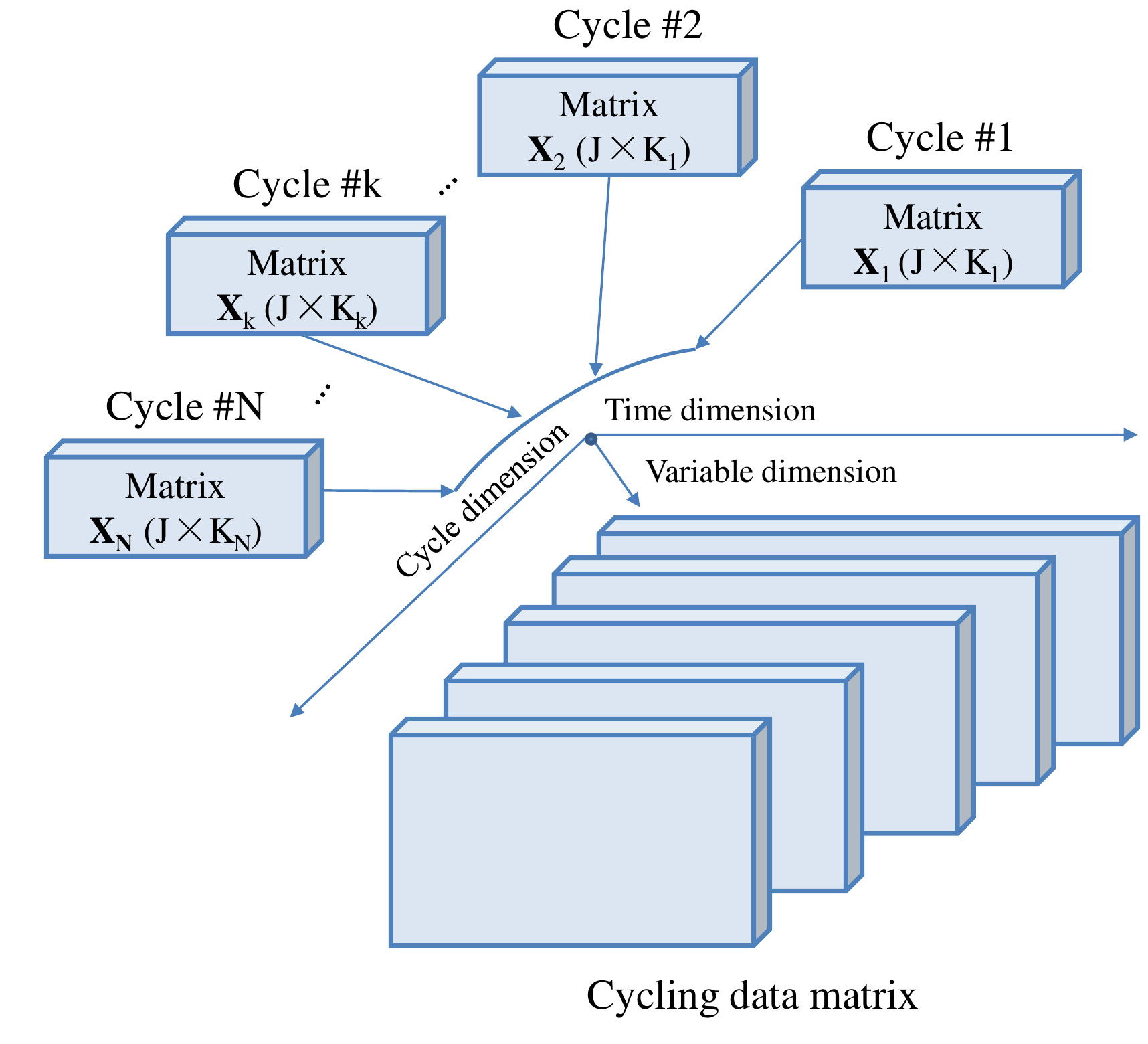}
\caption{Cycling discharging data and its data structure.}
\label{MyFig2}
\end{figure}

\section{Methodology}
Combined with cycling data characteristics, this section begins with the basic reorganizations about multi-stage division. Next, the three main parts of the proposed method are sequentially presented. First, dynamics hidden behind process variables are unfolded in the framework of phase subspace, yielding reconstructed data. Next, invariants indicating degradation trend are decomposed by linear superposition of the reconstructed data, which are computed using cycling SSA. Third, the monitoring strategy on invariants is given, distinguishing degradation stages from each other. The general flow of the proposed method is briefly illustrated in Fig. 3.

Since capacity continuously decreases over cycles, it is challenging to identify switch point, where two adjacent degradation stages can be clearly distinguished. Considering that capacity fading reflects the internal dynamics of measurements (i.e., voltage, current, and temperature in LiB), it is more reasonable to separate the whole degradation procedure according to the dynamical changes of these process variables. On the basis of this, the following considerations act as guidance:

(1) Despite the largely continuous nature of the capacity fading, it is possible that there are more than one degradation behaviors caused by complex external and internal factors;

(2) In each cycle that may exist features that are kept similar during the same degradation stage, while they are very dissimilar at different degradation stages.

With these considerations, the issue of identification degradation stages is equal to the problem of finding features that are sensitive to degradation changes. In this article, invariants across cycles are chosen as indicators and the specifics are illustration in the following parts.

\begin{figure}[htb]
\centering
\includegraphics[scale=0.30]{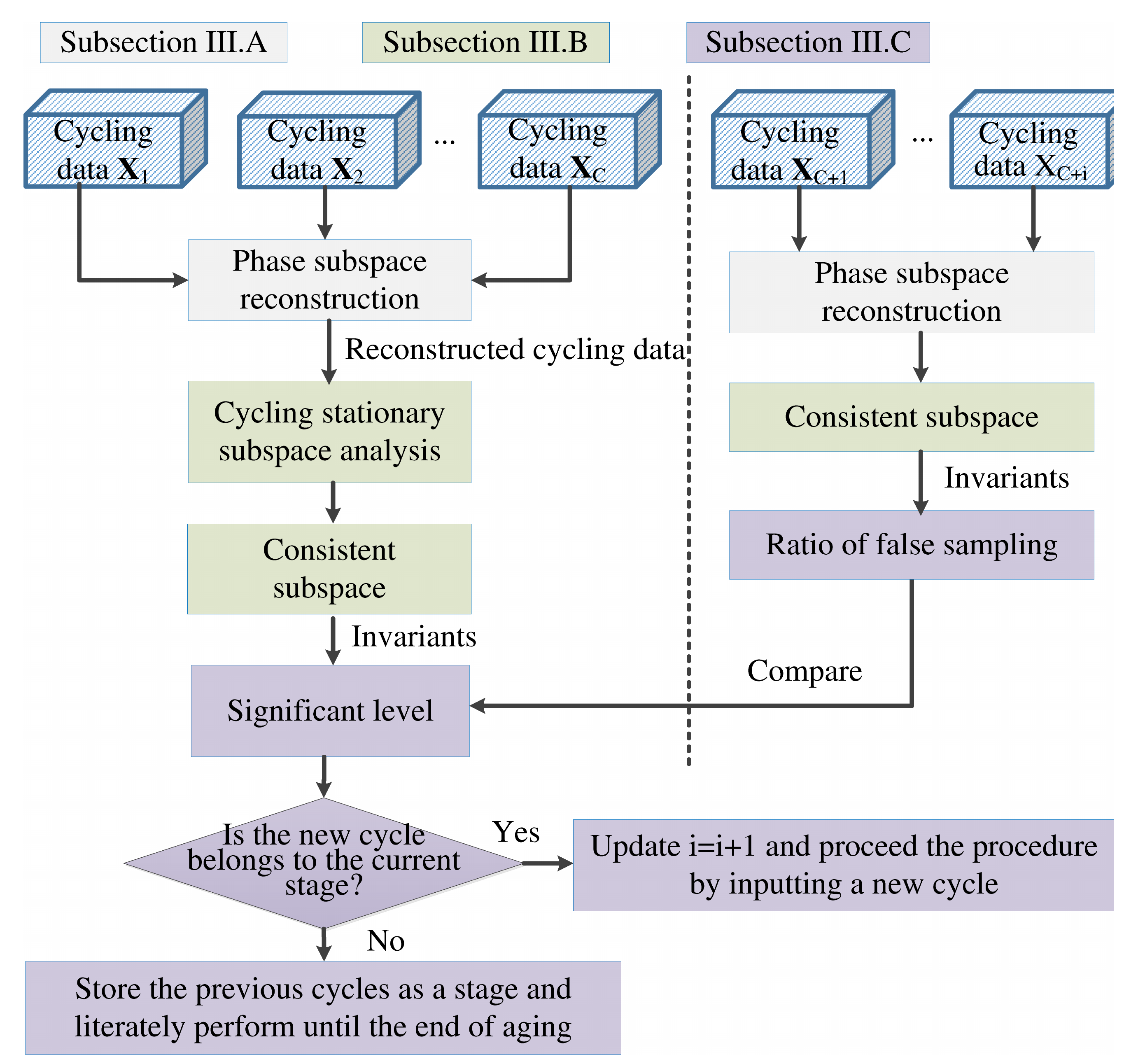}
\caption{Framework of the proposed method for multi-stage identification of battery performance degradation.}
\label{MyFig3}
\end{figure}

\subsection{Data Reconstruction in Multivariate Phase Subspace}
Phase subspace reconstruction (PSR) [9] originates from chaos theory, which aims to describe the underlying dynamics of a system by extending the low-dimension original measurements to a high-dimension embedding. This embedding ensures that hidden states are revealed through reconstruction without evolving of complex mechanisms. For a variable $\mathbf v=[v_1, v_2, ..., v_L]^T$ with the number of samples $L$, its reconstructed matrix is $\mathbf V=[\mathbf v_1, \mathbf v_2,..., \mathbf v_r]$ in phase space, which is achieved via delay stacked below,
\begin{equation}
\begin{aligned}
\mathbf{V}
&=\left[\mathbf{v}_{\mathbf{1}}, \mathbf{v}_{\mathbf{2}}, \ldots, \mathbf{v}_{r}\right] \\
&=\left\{\begin{array}{llll}
v_{1} & v_{1+\tau} & \cdots & v_{1+(r-1) \tau} \\
v_{2} & v_{2+\tau} & \cdots & v_{2+(r-1) \tau} \\
\cdots & \cdots & \cdots & \cdots \\
v_{k} & v_{k+\tau} & \cdots & v_{k+(r-1) \tau} \\
\cdots & \cdots & \cdots & \cdots \\
\end{array}\right\}
\end{aligned}
\end{equation}
where $\tau$ is time lag and $r$ is the embedding dimension.

$r$ determines the dimensions of reconstructed space, which can be computed using the false neighbor method [10]. A proper $\tau$ is determined when mutual information between $\mathbf x$, and its lagged vector is minimal [11]. More specifics about the calculation of $\tau$ and $r$ can be found in [12].

For multivariate time series with variable dimension $J$, reconstructed data matrixes are obtained by performing PSR on each variable. Noting the reconstructed matrixes as $\mathbf V_1, \mathbf V_2, ..., \mathbf V_J$ corresponding to each variable, respectively. Since the varying values of $\tau$ and $r$ of each variable, the dimensions of $\mathbf V_1, \mathbf V_2, ..., \mathbf V_J$ may be different, resulting a difficulty to merge these matrixes directly. Here, the maximum of $\tau$ among variables is employed to keep low correlations. And the maximum of $r$ among variables is selected to extend the variable dimension as many as possible. The integrated matrix with variable dimension $Jr$ is noted as $\mathbf V=[\mathbf V_1, \mathbf V_2, ..., \mathbf V_J]$.

\subsection{Learning Invariant with Cycling SSA}
Although many practical systems (e.g., stock market, geophysics) appear seriously non-stationary, they are possibly the superposition of stationary and non-stationary sources. That is, there may be invariant in dynamical system, which is the behind idea of SSA [13]. Motivated by this, SSA is employed to learn invariants from LiB discharging procedure, which possesses the non-stationary characteristics.

\subsubsection{Brief review of SSA}
For a set of non-stationary multivariate time series $\mathbf V$, it can be presented by a linear combinations of $d$ stationary source signals $\mathbf S_s=[\mathbf s_1, \mathbf s_2,..., \mathbf s_d]$ and $Jr-d$ non-stationary source signals $\mathbf S_{ns}=[\mathbf s_{d+1}, \mathbf s_{d+2},..., \mathbf s_J]$ as follows,
\begin{equation}
{\mathbf V}^T= \mathbf A \mathbf S=\left[\mathbf A_s  \mathbf A_{ns}\right]\left[\begin{array}{l}
\mathbf S_{s}^T \\
\mathbf S_{ns}^T
\end{array}\right]
\label{Eq1}
\end{equation}
where $\mathbf A$ is an invertible matrix with $Jr \times Jr$ dimensions; $\mathbf S$ is the collection of $\mathbf S_s$ and $\mathbf S_{ns}$, i.e., $\mathbf S=[\mathbf S_s  \mathbf S_{ns}]^T$; $\mathbf A_s$ is the loading matrix corresponding to $\mathbf S_s$ with $Jr \times d$ dimensions; $\mathbf A_{ns}$ is the loading matrix corresponding to $\mathbf S_{ns}$ with $Jr \times (Jr-d)$ dimensions.

Assuming $\mathbf A$ is invertible, $\mathbf S$ can be rewritten below,
\begin{equation}
\mathbf{S}=\mathbf{A}^{-1} \mathbf{V}^T=\mathbf{B W} \mathbf{V}^T
\label{Eq2}
\end{equation}
where $\mathbf{A}^{-1}$ is the invertible matrix of $\mathbf A$; $\mathbf W$ is a whitening matrix which is calculated as $\sqrt{\mathbf {V V}}^T$, and $\mathbf B$ is the projection matrix to be solved.

To solve stationary source $\mathbf S_s$, Kullback-Leibler divergence (KLD) [14] is employed to measure the distributions between $\mathbf S_s$ and Gaussian distribution with zero mean and unit variance. Correspondingly, $\mathbf B_s$ corresponding to minimal KLD is the optimal. In this way, the purpose of solving stationary source equals to finding the optimal $\mathbf B_s$.
\subsubsection{Cycling SSA for learning invariants}
The input of traditional SSA is continuous multivariate time series, which needs to be equally divided into several epochs to derive $\mathbf B_s$. To meet with the cycling characteristic of LiB data, a cycling SSA with PSR is specified below:

\textit {Step 1: Data reconstruction}

Taking $C$ discharging cycles as input data. Performing multivariate PSR given in Subsection III.A to each cycling data. Denoting the reconstructed data as $\mathbf X_1$, $\mathbf X_2$, to $\mathbf X_C$ corresponding to Cycles 1 to $C$.

\textit {Step 2: Data synchronization}

To meet with the precondition of equal length of epoches, a minimum length $M$ among all cycles is selected. By remaining the first $M$ samples in each matrix, the cycling data are synchronized to the same length. For brevity, the synchronized matrixes are still denoted as $\mathbf X_1$, $\mathbf X_2$, ..., $\mathbf X_C$.

\textit {Step 3: Problem formulation based on KLD}

The mean and variance of $\mathbf X_i (i \in[1, C])$ can be easily estimated as $\mathbf u_i$ and $\mathbf \Sigma_i$, respectively. Assuming $\mathbf B$ is available, perform linear transformation $\mathbf {s=BW} \mathbf X^T$ on each cycle. The mean and variance of the $i$th cycle after transformation are $\mathbf u_{i,s}=\mathbf {I_d} \mathbf {BW} \mathbf u_i$ and $\mathbf \Sigma_{i,s}=\mathbf I_d \mathbf {BW \Sigma_i} (\mathbf I_d \mathbf{BW})^T$, where $\mathbf I_d$ is the identity matrix $\mathbf I(J \times J)$ truncated to the first $d$ rows. In this way, solving Eq. (3) is transformed to an optimization problem by minimizing the objective function below,
\begin{equation}
\begin{aligned}
L(\mathbf{B}) &=\sum_{i=1}^{C} D_{KL}\left[N\left(\tilde{\mathbf{u}}_{i,s}, \tilde{\mathbf{\Sigma}}_{i, s}\right) \| N(0, \mathbf{I})\right] \\
&=\sum_{i=1}^{C} \sum_{m} P_{i,s}(m) \log \left(\frac{P_{i,s}(m)}{Q(m)}\right)
\end{aligned}
\end{equation}
where $\tilde{\mathbf{u}}_{i,s}=\mathbf I_d \mathbf {BW} \mathbf u_i$ and $\tilde{\mathbf{\Sigma}}_{i, s}= \mathbf I_d \mathbf {BW} \mathbf \Sigma_i (\mathbf I_d \mathbf {PW})^T$ are the mean and variance after linear transformation, respectively; $D_{KL}()$ is the KLD function; $N(0, \mathbf I)$ is the Gaussian distribution with zero mean and unit variance; $P_{i,s}(m)$ is the probability density of $N(\tilde{\mathbf{u}}_{i,s} ,\mathbf \Sigma_{i,s})$ for the $m^{th}$ sample; $Q(m)$ is the probability density of $N(0, \mathbf I)$ for the $m$th sample.

\textit {Step 4: Calculation of invariants}

According to [13], the optimal matrix $\mathbf B$ can be solved by conjugate gradient descend. Consequently, stationary sources are derived with the first $d$ rows of $\mathbf B$. The stationary space is the first $d$ dimension of matrix $\mathbf B$ denoted as $\mathbf B_s$. Then the invariants $\mathbf T_i$ in Cycle $i$ can be estimated as follows,
\begin{equation}
\mathbf{T}_{i}=\mathbf{I}_{d} \mathbf{B X}_{i}^T=\mathbf{B}_{s} \mathbf{X}_{i}^T
\end{equation}

\subsection{Identification Strategy of Multi-stages with Invariants}
Invariant has been learned in Subsections III.A and III.B, serving as data unit in the following identification strategy. The switch point of different stages is determined by checking the statistical consistency of invariants along cycle direction one by one. The specifics of the division procedure are given below:

\textit {Step 1: Data preparation}

From the beginning of LiB discharging process, prepare the reconstructed data matrixes $\mathbf X_1$, $\mathbf X_2$, to $\mathbf X_C$ according to Steps 1 and 2 of cycling SSA given in Subsection III.B.

\textit {Step 2: Invariant learning}

Calculate embedding space $\mathbf B_s$ among $\mathbf X_1$, $\mathbf X_2$, ..., $\mathbf X_C$ according to the Steps 3 and 4 of cycling SSA given in Section III. C. Then invariants $\mathbf T_i = \mathbf B_s \mathbf X_i i \in [1,C]$  are calculated by projecting $\mathbf X_i$ on $\mathbf B_s$.

\textit {Step 3: Threshold of normal invariants}

An extended multivariate time series $\mathbf T_s=[\mathbf T_1,\mathbf T_2,..., \mathbf T_C]$ is constructed by concatenating invariants in Step 2. However, it should be noted that the projection directions in $\mathbf B$ maybe correlated. That is, vectors in $\mathbf T_s$ may not be orthogonal to each other, resulting in redundant information. Principal component analysis [14] is performed on $\mathbf T_s$ to approach the orthogonal latent variables below,
\begin{equation}
\overline{\mathbf{T}}_s = {\mathbf{P}_s \mathbf{T}_{s}}
\end{equation}
where $\overline{\mathbf{T}}_s$ is the principal component (PC) matrix; $\mathbf P_s$ is the stage loading matrix retaining all PCs.

In fact, most of process variations in $\mathbf T_s$ are contained in the first several PCs. The retained number of PCs $R$ can be determined by the cumulative explained variance rate [14] to keep most of the process variability (85$\%$ here). Monitoring statistics Hoteling-$T^2$ (threshold) is designed to monitor the normal variation of $\mathbf T_s$, which can be approximated by an $F$ distribution [14] as follows,
\begin{equation}
T_{s}^{2}=\overline{\mathbf{T}}_{s,R} \mathbf \Xi_{s}^{-1} \overline{\mathbf{T}}_{s,R} \sim \frac{R\left(N_C^{2}-1\right)}{N_C(N_C-1)} F_{R, N_C-R, \alpha_{s}}
\end{equation}
where $\overline{\mathbf{T}}_{s,R}$ is the first $R$ rows of $\overline{\mathbf{T}}_{s}$ in Eq. (6), $\mathbf \Xi_s$ is the covariance matrix of $\overline{\mathbf{T}}_{s,R}$, $\alpha_s$ is the significant level (here is 0.05) to derive the 95$\%$ confidence limit, and $N_C$ is the total number of samples.

\textit {Step 4:  Invariant projection for following cycles}

For the $(C+1)$th cycle, data matrix $\mathbf X_{C+1}$ is reconstructed with the same processing as shown Step 1. Projecting $\mathbf X_{C+1}$ on embedding $\mathbf B_s$ obtained in Step 2 and the corresponding invariants are calculated as $\mathbf T_{C+1}= \mathbf B_s \mathbf X_{C+1}$. Further, the orthogonal invariants after removing redundancy is $\overline {\mathbf T}_{C+1,R}= \mathbf P_{s,R} \mathbf T_{C+1}$, where $\mathbf P_{s,R}$ is the first $R$ rows of $\mathbf P_s$ in Eq. (6). After that, monitoring statistics $t_{C+1}^{2}(k)$ at sampling time $k$ in this cycle is calculated as,
\begin{equation}
t_{C+1}^{2}(k)=\overline{{T}}_{C+1,R}(k)^{T} \mathbf \Xi_{s}^{-1} \overline{{T}}(k)_{C+1,R}
\label{Eq7}
\end{equation}
where $\overline{\mathbf{T}}_{C+1,R}(k)$ is the $k$th sample in $\overline{\mathbf{T}}_{C+1,R}$.

\textit {Step 5: Comparison of statistics with its control limits}

The index abnormality rate $AR_{C+1}$ is defined as the ratio that the number of abnormal samplings over the total samplings in Cycle ($C+1$). The abnormal sampling interval is identified where $t_{C+1}(k)^{2}>T_{s}^{2}$. The switch point is where two adjacent stages can be identified through continuously comparing $AR_{C+1}$ with its significant level, which is $\alpha_s$ as given in Eq. (7):

Case I: If $AR_{C+1}<0.05$, it means that invariants in this cycle are consistent with that of the previous cycles. In this case, Cycle $C+1$ belongs to the same degradation stage. Then update $C=C+1$ and go to Step 4.

Case II: If $AR_{C+1}>0.05$  and $AR_{C+2}>0.05$  hold for consecutive two cycles, it indicates the obvious changes of invariants. A new degradation stage begins from Cycle $C+1$ since invariants are over its normal region. Then update $C=C+1$ and go to Step 6.

\textit {Step 6: Data updating and recursive implementation}

Remove the identified degradation stage and the left cycling data are now employed as the new input data in Step 1. Recursively repeat Steps 2-6 from the updated beginning of the LiB discharging process to find the remaining stages.

From Steps 1 through 6, the output of the proposed method is a partition of degradation procedure along cycle direction. The behind idea is to sequential cluster those cycles with consistent invariants so that they can be presented by the same degradation behavior, while those different ones will be classified into different degradation stages.


\section{Illustration and Discussion}
In this section, the proposed multi-stage identification approach is verified with a well-known benchmark for LiB study. This benchmark is released by NASA . Four battery datasets that have been widely used in many previous researches [2]-[4] are employed. The specifics of these datasets are described in Table I. Due to page limitation, only B0005 is carefully analyzed to illustrate performances with respect to data reconstruction, invariant learning, and the degradation stage division. Besides, division results of degradation stage for other batteries are shown.

\begin{table}[htb]
\scriptsize
\renewcommand{\arraystretch}{1}
\caption{Information of the employed datasets}
\label{table_example}
\begin{center}
\begin{threeparttable}
\begin{tabular}{c c c c}
\hline
\toprule  
{Name} & {Rated capacity (Ah)} & {No. of cycles} & {Measured variables} \\
\hline
B0005 & 2 & 166 & Current, voltage, temp. \\
\hline
B0006 & 2 & 167 & Current, voltage, temp. \\
\hline
B0007 & 2 & 167 & Current, voltage, temp. \\
\hline
B0018 & 2 & 132 & Current, voltage, temp. \\
\bottomrule 
\end{tabular}
\end{threeparttable}
\end{center}
\end{table}

\subsection{Multivariate PSR Reconstruction}
From beginning of the discharging process, the first fifteen cycling data are used for PSR, denoted as $\mathbf X_1$, $\mathbf X_2$, ..., $\mathbf X_{15}$. According to the selection strategy given in Section III. A, tunable parameters $\tau$ and $r$ are determined as 5 and 3, respectively. Fig. 4(a) shows the three-dimension projection in phase space of a voltage signal for the fifteen cycling data. Although the varying trend in original measurements is very similar across cycles in Fig. 4 (b), the slow shift can be observed from the reconstructed cycling data. That is, more dynamics about the variability along cycles are revealed in the reconstructed data. Moreover, PSR provides a way to visualize this variability.
\begin{figure}
  \centering
  \includegraphics[scale=0.7]{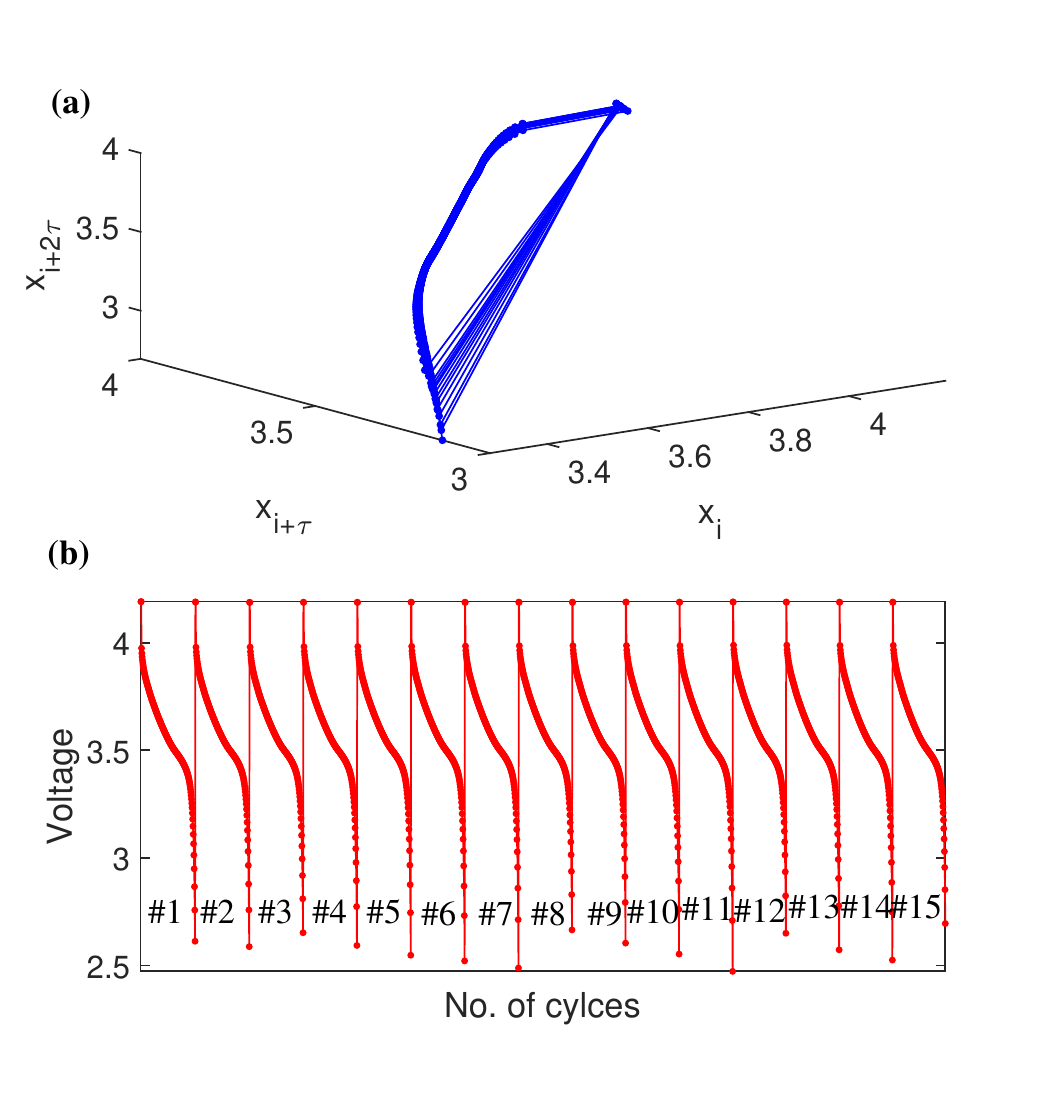}\\
  \caption{(a) Reconstructed data in PSR of cycling voltage signal from Battery B0005 (b) Original time-series .}
  \label{MyFig4}
\end{figure}

\subsection{Invariants using Cycling SSA}
With the reconstructed data in the last section, these cycling data are synchronized to the same length, which is 165 here. According to the selection strategy given in Section III.B, three stationary sources are retained using ADF test, which could indicate the stationary of a variable by judging the existence of unit root. Fig. 5 visualizes these stationary sources, which present invariant across cycles. This proves that inferences that there may be consistent dynamics over cycles.

\begin{figure}
  \centering
  \includegraphics[scale=0.55]{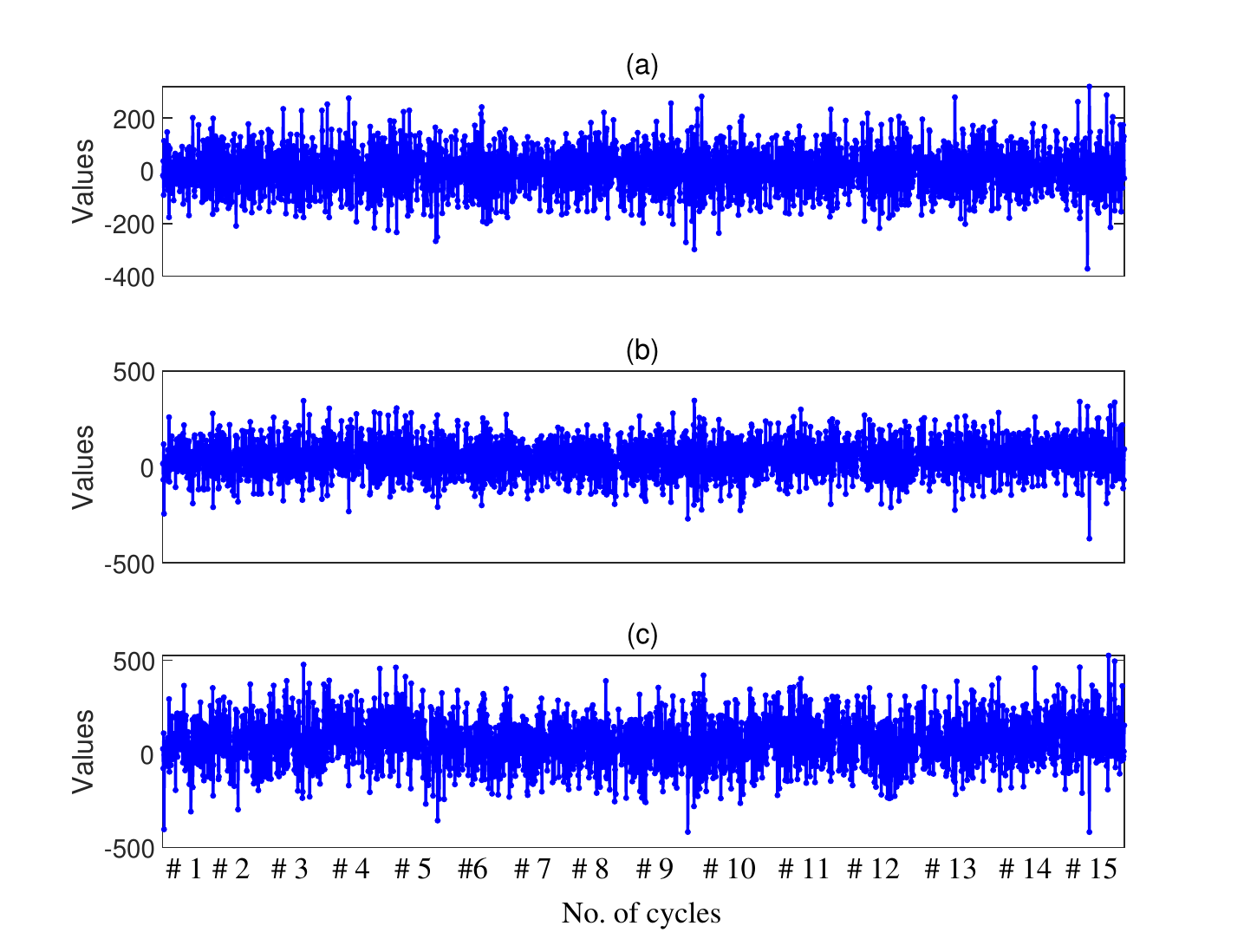}\\
  \caption{(a) the first stationary source (b) the second stationary source, and (c) the third stationary source decomposed from B0005 based on cycling SSA.}
  \label{MyFig5}
\end{figure}

\subsection{Identification of Degradation Stages}
Based on the extracted invariants from the first fifteen cycles, control limit $T^2$ defined in Eq. (7) is calculated with the significance level of 0.05. Iteratively perform cycling SSA method on the remaining cycling data. Based on the obtained stationary subspace, invariants in these cycles are obtained. In Fig. 6, different degradation behaviors are observed at the 27th cycle, where the rate of false classification rises rapidly and above the normal level ever since. Therefore, the cycles from 1 to 26 are regarded have consistent invariant and they form the first degradation stage.
\begin{figure}
  \centering
  \includegraphics[scale=0.5]{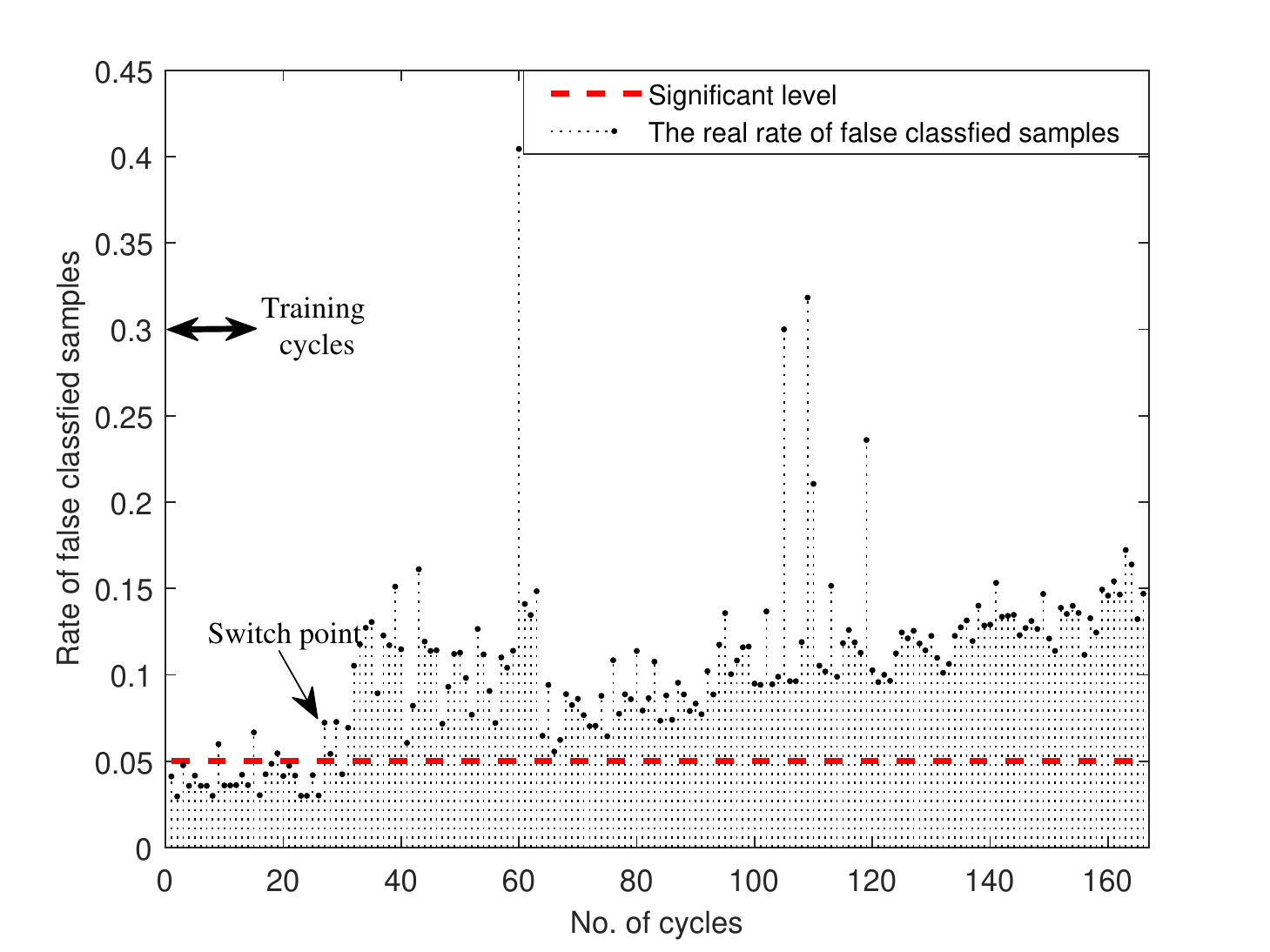}\\
  \caption{Phase division of the first stage for Battery B0005.}
  \label{MyFig6}
\end{figure}

Removing the first degradation stage, the remaining 140 cycles are used as updated input for further stage identification. First of all, the first fifteen cycling data (i.e., Cycle 27 to Cycle 41) are picked out for data reconstruction. Then invariants across these cycles are decomposed for development of normal control limit of the second degradation stage. By counting the abnormal samples over the control limits in each cycle, the false misclassification rate of each cycle will be calculated. Fig. 7 visualizes results and it is observed that the two consecutive cycles 83 and 84 show an abnormal behavior. According to judgment rules, Cycle 83 is regarded as a new switch point, which indicates the existence of the third degradation stage.

\begin{figure}
  \centering
  \includegraphics[scale=0.55]{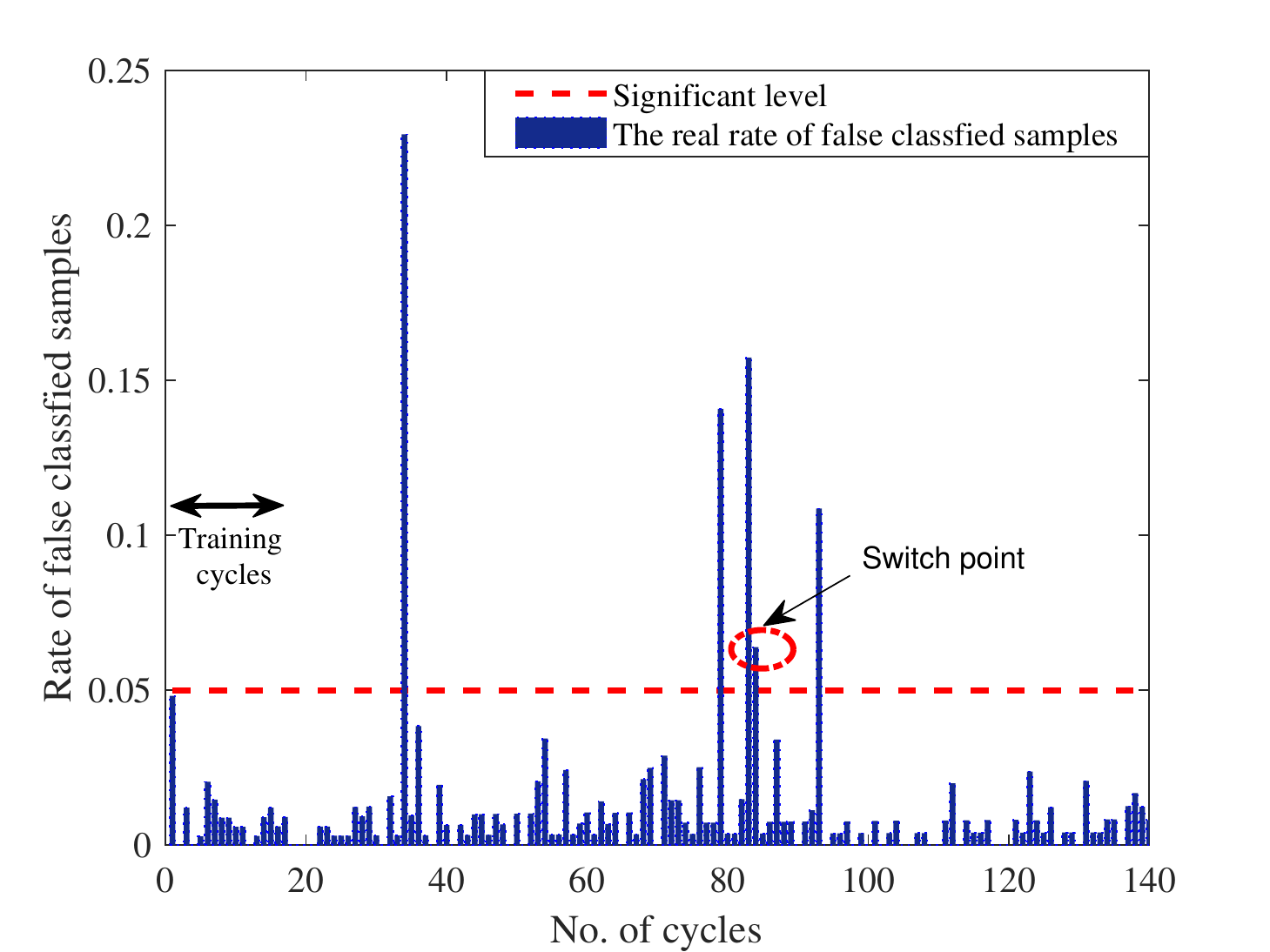}\\
  \caption{Phase division of the second stage for Battery B0005.}
  \label{MyFig7}
\end{figure}

Removing both the first stage (1-26) and the second stage (27-108), the remaining 58 discharging cycles are updated as input. With the similar procedures, no consecutive abnormal cycles are observed as shown in Fig. 8. It indicates that all cycles have consistent invariants. Combining with previous analysis, the whole degradation procedure is divided into three stages, in which different degradation behavior is identified.

\begin{figure}
  \centering
  \includegraphics[scale=0.55]{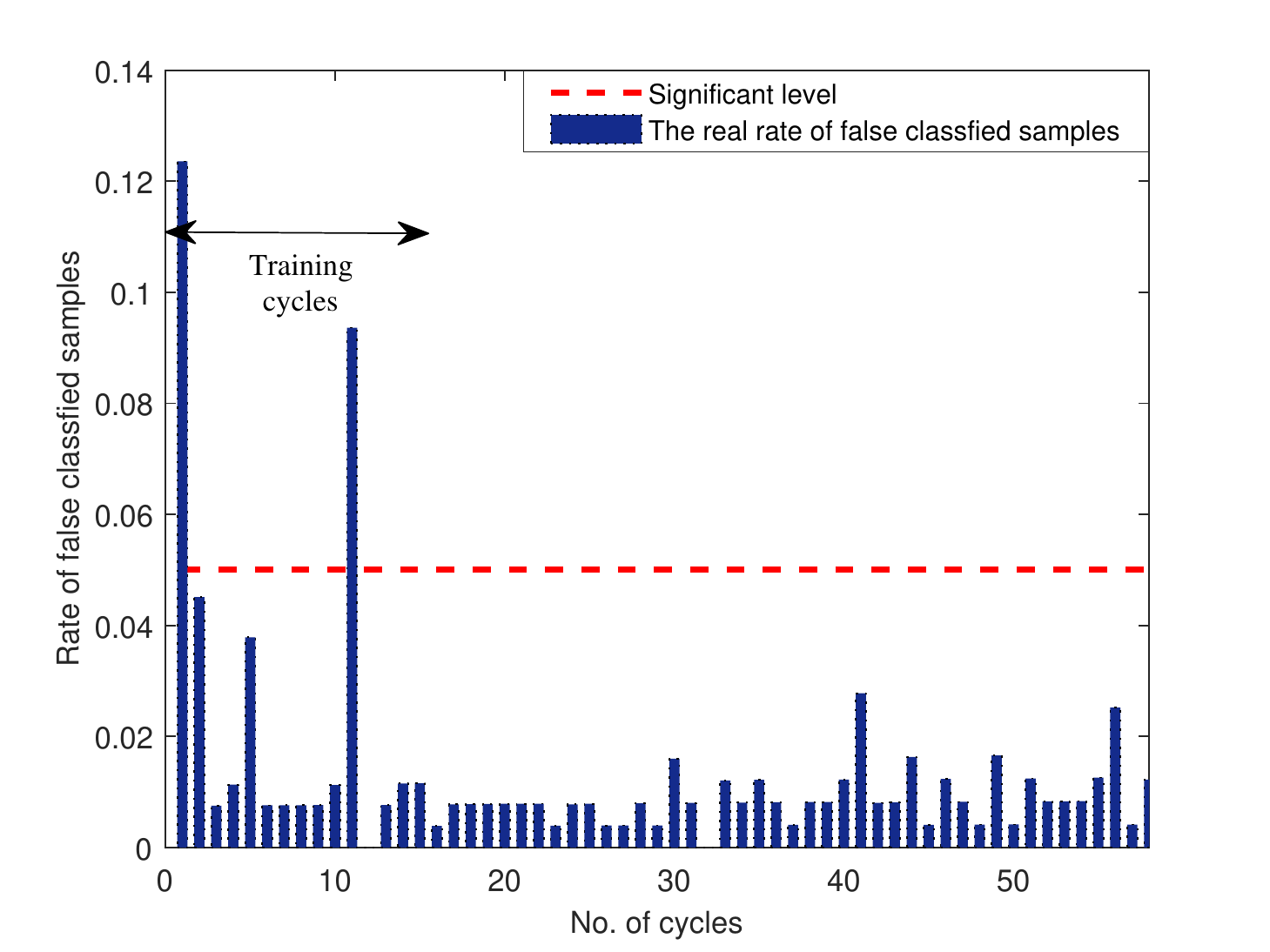}\\
  \caption{Phase division of the third stage for Battery B0005.}
  \label{MyFig8}
\end{figure}

In addition to battery B0005, another three batteries are analyzed according to the same procedure. The details are omitted for brevity. Table II summarizes the results, in which the start and end of each degradation stage have been specified. By reviewing the results, several phenomena can be easily observed. First, B0006 has a very similar evolving trend with B0005, presenting the number of stages and cycles. Second, the battery with the same material may have very different degradation behavior.

\begin{table}[htb]
\scriptsize
\renewcommand{\arraystretch}{1}
\caption{Multi-stage partition results for other three datasets}
\label{table_2}
\begin{center}
\begin{threeparttable}
\begin{tabular}{c c c}
\hline
\toprule  
{Name} & {No. of stages} & {Range}\\
\hline
B0006 & 3 & 1-26,27-110,111-167 \\
\hline
B0007 & 4 & 1-60,61-79,80-129,130-167 \\
\hline
B0018 & 2 & 1-53,54-132 \\
\bottomrule 
\end{tabular}
\end{threeparttable}
\end{center}
\end{table}

As the first attempt in this field, comparisons with previous work about identification accuracy is not available. Alternatively, we focus on illustrating the efficacy of the proposed algorithm by in-depth process understanding.
\section{Conclusion}
To probe into the degradation mechanism of LiB from data-driven perspective, this article puts forward a multi-stage division for the first time in improving process insights. In-depth analysis is achieved through three main parts, including data reconstruction in phase subspace, invariant learning with improved cycling stationary subspace analysis, and a stage division strategy. According to the simulation results, implicit dynamics could be unfolded with data reconstruction, indicating the evolution of sequential cycling data. Invariants could be effectively learned and used for indication of degradation switch. Additionally, it would be interesting to apply this idea for other topics, such as state of health estimation, and better performances may be achieved.

\section{Acknowledgement}
This work was supported by the National Natural Science Foundation of
China (No. 61903327).

\vspace{12pt}
\end{document}